\documentstyle[prb,aps,preprint,epsf]{revtex}
\begin{document}

\title{ {\bf Effect of Defects on the Line shape of Electron
Paramagnetic Resonance Signals from the Single-Molecule Magnet
Mn$_{\mathbf 12}$}: A Theoretical Study}
\draft
\author{Kyungwha ${\mathrm Park^{1,2,\ast}}$ 
\and N.~S.\ ${\mathrm Dalal^{2,\dag}}$, 
and P.~A.\ ${\mathrm Rikvold^{1,3,\ddag}}$}
\address{${\mathrm ^{1}}$School of Computational Science and 
Information Technology, Florida State University, Tallahassee, 
Florida 32306-4120 \\
${\mathrm ^{2}}$Department of Chemistry, Florida State University,
Tallahassee, Florida 32306-4390 \\
${\mathrm ^{3}}$Center for Materials Research and 
Technology and Department of Physics, 
Florida State University, Tallahassee, Florida 32306-4350}
\date{\today}
 
\maketitle
\newpage

\begin{abstract}
We herein estimate the effect of lattice defects on the line shape of 
electron paramagnetic resonance (EPR) signals from a single crystal of
the $S=10$ single-molecule magnet Mn$_{12}$ with the external 
magnetic field along the crystal $c$ axis. A second-order perturbation 
treatment of an effective single-spin Hamiltonian indicates that a small, 
random, static misorientation of the magnetic symmetry axes in a crystalline 
lattice can lead to asymmetric EPR peaks. Full spectra are simulated
by calculating probability-distribution functions for the resonant fields, 
employing distributions in the tilt angle of the easy axis from the
$c$ axis, in the uniaxial anisotropy parameter, and in the $g$-factor. 
We discuss conditions under which the asymmetry in the EPR spectra 
becomes prominent. The direction and magnitude of the asymmetry 
provide information on the specific energy levels involved with 
the EPR transition, the EPR frequency, and the distribution 
in the tilt angle.
\end{abstract}


\section{Introduction}

\begin{figure}
\begin{center}
\epsfxsize=7.5cm
\epsfysize=7.5cm
\epsfbox{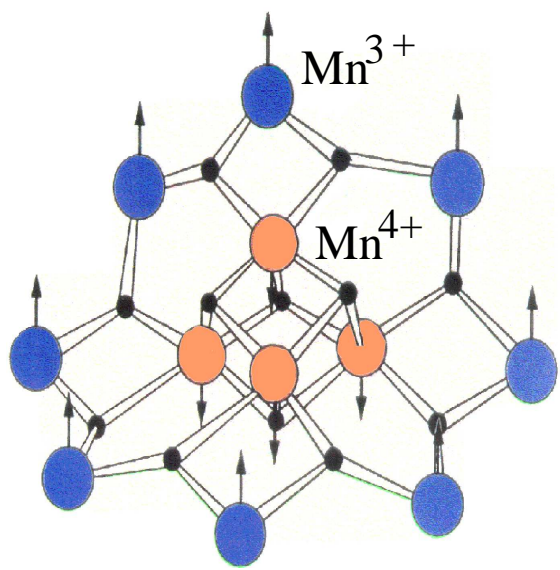}
\vspace{0.1in}
\caption{Schematic diagram of the magnetic core of 
the Mn$_{12}$ molecule. Each molecule includes
a tetrahedron of four  Mn$^{4+}$($S=3/2$) ions at the center,
surrounded by eight Mn$^{3+}$($S=2$) ions, yielding
an effective ground-state spin of $S=10$
($S=8\times 2 - 4 \times 3/2$). Each molecule has a tetragonal
symmetry.}
\label{structure}
\end{center}
\end{figure}

The nanomagnetic compound [Mn$_{12}$O$_{12}$(CH$_3$COO)
$_{16}$(H$_2$O)$_4$]$\cdot$2CH$_3$COOH$\cdot$4H$_2$O
(abbreviated hereafter as Mn$_{12}$),\cite{LIS80} first synthesized
by Lis,\cite{LIS80} consists of a core of twelve Mn ions,
linked together by oxygen atoms from the bridging acetate ligands.
The Mn ions form the magnetic core, and are shown schematically
in Fig. \ref{structure}. The Mn ions are of mixed valence: eight
of them (Mn$^{3+}$) with four unpaired electrons ($S=2$) form
the outer ring, and are ferromagnetically coupled within
themselves, to lead to a total spin of $S_1=8\times 2=16$. The other
four ions (Mn$^{4+}$) each with $S=3/2$ form the center of the core
and are arranged in a Mn-O cubane type structure. These are also
ferromagnetically coupled within themselves, with a total spin of
$S_2=4\times (3/2)=6$. The total spin on a Mn$_{12}$ molecule has been
determined to be 10, which can be understood as arising from
a strong antiferromagnetic coupling between the two $S_1$ and $S_2$
sub-systems.\cite{SESS93-1,ROBI00,GOTO00} 
The whole molecule may thus be considered as a $S=10$
particle as far as the ground state is concerned, which is of 
primary interest in the present work. The acetate ligands are
covalently bonded via the oxygens to the Mn ions.
There are two acetic acid molecules and four H$_2$ molecules solvated 
in the lattice. The magnetic moments of the Mn$_{12}$ molecules are 
most easily aligned along the crystal $c$ axis, which is thus 
the easy axis for a perfect crystal (without imperfections/dislocations).
There is a zero-field energy barrier of approximately 65 K 
against magnetization reversal.\cite{BARR97,MIRE99}
The effective distance between different Mn$_{12}$ molecules is
approximately 14~\AA.\cite{LIS80,PARK02-1,PARK02-2}

Earlier, we have examined the effect of possible defects on the linewidths
of electron paramagnetic resonance (EPR) spectra for the single-molecule
magnet Mn$_{12}$,\cite{PARK02-1,PARK02-2} and have compared our calculated
results with experiments.\cite{PARK02-1,HILL02} 
The imperfections in large complex molecules 
such as Mn$_{12}$ could be of many types, such as 
dislocations\cite{CHUD01} and the disorder
caused by the various orientations of the acetate ligand and
of the solvent molecules in a unit cell.\cite{CORN01} 
So far the origin of the possible defects has not yet been clarified.
In the study,\cite{PARK02-1,PARK02-2} we modeled that the defects could be 
represented by the distributions in the uniaxial anisotropy parameter 
$D$ (second-order anisotropy) and the $g$-factor. To calculate the
EPR linewidths, the intermolecular dipolar interactions and 
the distributions in $D$ and $g$ were considered. For Mn$_{12}$
the intermolecular dipolar interactions ($\sim$ a couple of
hundred gauss) are substantially smaller than the the effect 
of the distributions in $D$ and $g$ ($\sim$ a thousand gauss). 
The detailed estimates of the dipolar interactions and the effect of the
distributions in $D$ and $g$, were given in Refs.~\cite{PARK02-1,PARK02-2}
as a function of temperature, resonance frequency, and the energy
levels relevant to the EPR transition. Our calculated results
\cite{PARK02-1,PARK02-2} agree well with experiments.\cite{PARK02-1,HILL02}

In the present study, we describe our theoretical investigation 
of the effects of lattice defects on the signal line {\it shape} 
of EPR spectra for Mn$_{12}$. We propose that the lattice defects 
may cause a small, static, random misorientation of the magnetic
symmetry axes of each molecule in the lattice. Here we examine 
the full microwave absorption line shape as a function of
resonance frequency and the energy levels relevant to the EPR 
transitions, including Gaussian distributions
in the tilt-angle (the angle between the magnetization easy axis and the 
crystal $c$ axis) as well as the distributions in $D$ and $g$.
Then we find that consideration of the Gaussian distribution in 
the tilt angle leads to a line-shape asymmetry in the spectra. 
Conversely, it might be possible to learn about the defect 
structure from this asymmetry.

This study was prompted by three factors. First, Mn$_{12}$
and its analogs, with their high-spin ground states of $S=10$, exhibit
novel properties, such as macroscopic quantum tunneling 
(MQT),\cite{GUNT95,CHUD98} whose origin has not yet been fully 
understood. A new model of MQT, especially for Mn$_{12}$, 
involves a direct role for lattice defects.\cite{CHUD01}
Second, Mn$_{12}$ has recently been proposed as a material 
for potential use in quantum computation.\cite{LOSS01} For this
application, a detailed knowledge of the line shape of the EPR absorption
spectrum would be required. The proposed method can be applied to
microwave transitions between the various spin
levels. Third, many earlier high-field EPR studies of Mn$_{12}$ and 
related systems have reported asymmetric line shapes, but the origin of the
asymmetry has not been clarified.\cite{BARR97,BARR00} 
It was assumed that the asymmetry
might be an instrumental artifact, resulting from the existence of
standing waves in the waveguides.\cite{BARR97,BARR00} 
 
Asymmetric EPR line shapes have been known since the
early days of EPR spectroscopy. For example, Bleaney and Rubins 
considered forbidden hyperfine transitions and noted 
that EPR line shapes from powder samples, which have
a distribution of particle orientations,
can be asymmetric.\cite{BLEA61} Although we use a standard
spin Hamiltonian [Eq.~(\ref{eq:ham_hz_xyz})] and 
routine perturbation technique, our study is distinct 
from such earlier studies in the following ways:
(a) We consider single-crystal samples, with
measurements along a well-defined direction such as
the crystal $c$ axis. (b) Our system involves the
existence of strong intramolecular spin-exchange effects within the Mn$_{12}$
molecule (which is an order of 100 K),\cite{XXXXX} 
which obliterates the hyperfine structure and the effects of any
forbidden transitions. (c) The system and experimental situation are
characterized by a dominant uniaxial anisotropy term in the
spin Hamiltonian and a strong external magnetic field, 
and are thus quite applicable to high-field EPR.
(d) We examine systematically the asymmetry in the EPR line shape 
as a function of the energy levels between which the microwave transition
occurs, the microwave frequency, and the distribution in the
tilt angle. 
Future experimental tests of our predictions for the EPR line shapes of
single-molecule magnets should be fruitful. 

The remainder of this paper is organized as follows.
In Sec.~II the model is described, and in Sec.~III the resonant
field is calculated as a function of the rotation angle $\theta$
(which is the tilt angle of the molecular easy axis with respect to the 
crystal $c$ axis), and the direction and the magnitude of the asymmetry 
in the resonant fields are discussed. In Sec.~IV we calculate 
the probability-distribution functions of the resonant fields,
including the distributions in 
$\theta$, $D$, and $g$, to predict the asymmetry in the spectra. 
In Sec.~V we present a discussion and our conclusions.

\section{Model}

In the presence of defects or impurities in a sample,
Mn$_{12}$ molecules can be displaced or rotated from their normal positions. 
Consequently, each molecule sees a slightly different
crystal field caused by the surrounding molecules, compared to that
seen in a perfect crystal. We propose that this slightly 
different crystal field seen by a molecule results in a small, 
static, random misorientation (rotation) of the magnetic symmetry
axes of the molecule in the crystalline lattice. The degree of
the misorientation can be quantified as the magnetization easy axis 
of the molecule is rotated by an angle $\theta$ 
away from the crystal $c$ axis. The majority of the molecules 
are assumed to have their easy axes almost aligned with the $c$ axis, 
so the tilt angle $\theta$ is assumed to have a Gaussian distribution 
about zero with a small standard deviation. 
Hereafter $a$, $b$, and $c$ denote the crystal axes, 
while $x$, $y$, and $z$ denote the molecular 
(magnetic) anisotropy axes of a single molecule. In this study, 
we consider the case of varying the magnetic field at a fixed 
EPR frequency (roughly 50 to 500 GHz) when the external magnetic field 
is applied along the $c$ axis.

We confine our study to the uniaxial molecular magnet, Mn$_{12}$,
because of its very small transverse anisotropy 
compared to its uniaxial anisotropy.\cite{BARR97,PERE98} 
Additional transverse anisotropy terms or small transverse magnetic 
fields at $\theta=0$
may also produce asymmetry in the spectra, because of nonlinear relationships
between the energy cost of an EPR transition and the sweeping field.
For Mn$_{12}$, the negligibly small transverse anisotropy enables us to 
concentrate on the asymmetry effect 
caused by a distribution in $\theta$ only.

Our goal is to investigate how static, random rotations of the
magnetic anisotropy axes caused by defects affect 
the line shapes of EPR spectra as functions of frequency and energy level, 
and to predict EPR spectra which may be compared with
experimental data in near future.
For the single crystal of Mn$_{12}$ examined in
Refs. \onlinecite{PARK02-1,HILL02}, 
it was found that the effects of $D$-strain and $g$-strain are 
more significant than dipolar interactions.\cite{PARK02-1,PARK02-2}  
Therefore we need to combine the effects of $D$-strain and $g$-strain 
with the effect of static, random rotations 
of the easy axes in order to obtain realistic spectra.
After analytically calculating the resonant field as a function of 
$\theta$, $D$, and $g$, we numerically obtain the probability 
distribution function of the resonant field, assuming Gaussian 
distributions in $\theta$, $D$ and $g$. We do not take into 
account the effects of natural 
linewidths, of dipolar interactions between molecules, or of temperature
on the line shapes.

When an external magnetic field is applied along the $c$ axis, and the
magnetic anisotropy easy axis of a single molecule (the $z$ axis) is tilted 
by $\theta$ away from the $c$ axis,
the single-spin ground-state Hamiltonian is, to lowest order, 
in terms of the spin operators along the molecular axes,
\begin{eqnarray}
{\cal H}&=&-D S_z^2 - g \mu_B B \cos \theta~S_z 
- g \mu_B B \sin \theta (\sin \psi~S_x + \cos \psi~S_y) ~,
\label{eq:ham_hz_xyz} 
\end{eqnarray}
where $D=0.55$~$k_B$,\cite{BARR97} the isotropic $g=1.94$, \cite{BARR97}
$\mu_B$ is the Bohr magneton, and $B$ is the magnitude of the external field.
Here, $\psi$ is the third of the standard Euler angles $(\theta, \phi, \psi)$
as defined in Ref. \onlinecite{GOLD}. 
(For the rotation matrix, see Appendix A.)
There is no constraint on the magnitude of $\psi$ 
because it does not affect the energy eigenvalues of the spin Hamiltonian.

For simplicity, we ignore small
fourth-order anisotropy terms\cite{BARR97,MIRE99}
in the single-spin Hamiltonian. 
This is justifiable because: (i) these terms make rather small 
energy contributions to the eigenvalues, 
and (ii) some experimental data indicate that the magnitude 
of the transverse fourth-order terms for Mn$_{12}$ may be too 
small to be measured via, for example, neutron scattering.\cite{BAO00,MIRE00}

We can also write the single-spin Hamiltonian in terms of spin operators
along the {\it crystal} axes. Both forms of the spin Hamiltonian represent 
the same system. (See Appendix A.)

\section{Resonant fields}

\begin{figure}
\begin{center}
\epsfxsize=7.5cm
\epsfysize=7.0cm
\epsfbox{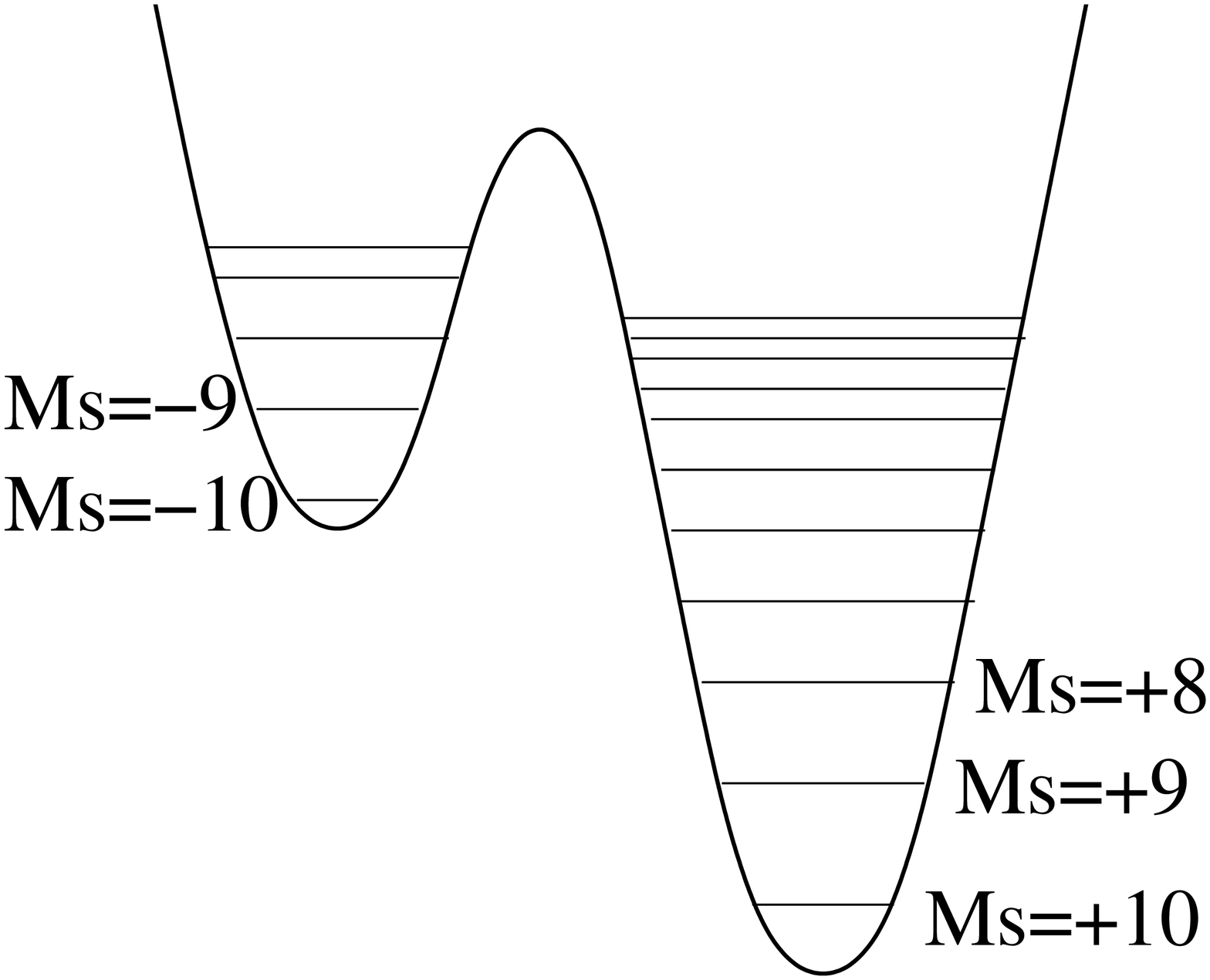}
\vspace{0.1in}
\caption{Schematic diagram of the potential-energy wells for 
the molecular magnet Mn$_{12}$,
which has an effective spin $S=10$. Here $M_s$ is the eigenvalue
of the spin operator $S_z$. For the
lower energy well (the right-hand well), the difference
between adjacent energy levels is $g \mu_B B + D(2M_s-1)$. For
the higher energy well (the left-hand well), the difference
is $-g \mu_B B + D(2M_s-1)$. As the applied field increases,
the difference between the energy levels $M_s=-10$ and $M_s=+10$ increases.}
\label{energywell}
\end{center}
\end{figure}

To calculate perturbatively, to lowest order, the resonant fields 
with the spin Hamiltonian (\ref{eq:ham_hz_xyz}), we assume that
the tilt angle $\theta$ is very small. This assumption 
is probably valid because a large misalignment between the crystal $c$ axis 
and the magnetic easy axis of a molecule (the $z$ axis) 
is not expected for Mn$_{12}$, since the reported X-ray
data\cite{XXX} on bond angles and distances are quite precise, with no
mention of mosaicity issues.
Then taking $V \equiv -g \mu_B B \sin\theta (\sin\psi~S_x + \cos\psi~S_y)$
as a small perturbation to the rest of the terms in ${\cal H}$,
we obtain the second-order perturbed energy 
${\mathcal E}^{(2)}_{M_s}$ of the level $M_s$, where
$M_s$ is an eigenvalue of the spin operator $S_z$:
\begin{eqnarray}
{\mathcal E}^{(2)}_{M_s}&=& \frac{\eta^2 (S-M_s)(S+M_s+1)}
{{\mathcal E}^{(0)}_{M_s}-{\mathcal E}^{(0)}_{M_s+1}}
+ \frac{\eta^2 (S+M_s)(S-M_s+1)}
{{\mathcal E}^{(0)}_{M_s}-{\mathcal E}^{(0)}_{M_s-1}} \;,
\label{eq:e2}
\end{eqnarray}
where $\eta \equiv (g \mu_B B \sin\theta)/2$, and 
${\mathcal E}^{(0)}_{M_s}=-D M_s^2 - g \mu_B B \cos\theta~M_s$ 
is the unperturbed energy
eigenvalue of the level $M_s$. With the selection rule that 
EPR transitions occur between adjacent energy levels only, 
we have at resonance
$h \nu={\mathcal E}^{(0)}_{M_s \pm 1}
+{\mathcal E}^{(2)}_{M_s \pm 1}
-{\mathcal E}^{(0)}_{M_s}-{\mathcal E}^{(2)}_{M_s}$,
where $\nu$ is the frequency of the applied microwave radiation
in the EPR experiment.
The resonance equation is quartic in $B$, so its solutions are
too complicated to provide useful information.
To obtain a simple analytic expression for the
resonant field, we substitute the unperturbed expression for
the resonant field $B_0$ [for example, for the transition between 
the levels $M_s$ and $M_s \pm $1, 
$B_0=(\mp h \nu - D(2M_s \pm 1))/g\mu_B \cos\theta$],
for $B$ in Eq.~(\ref{eq:e2}).
Then we obtain the approximate resonant fields as follows:
\begin{eqnarray}
B_{\mathrm res}(M_s \rightarrow M_s-1)\! \! &=& \! \!
B_0 (M_s \rightarrow M_s-1) - 
\frac{g \mu_B B_0^2 
\sin^2\theta~X(\nu, M_s)}{4 \cos\theta}, \: M_s=10,9,...
\label{eq:hres_hz_3} \\
B_{\mathrm res}(M_s \rightarrow M_s+1)\! \! &=& \! \!
B_0 (M_s \rightarrow M_s+1) + 
\frac{g \mu_B B_0^2 
\sin^2\theta~X(\nu, -M_s)}{4 \cos\theta}, \: M_s=-10,-9,...
\label{eq:hres_hz_4} \\
X(\nu, M_s)&=& \frac{2(S-M_s+1)(S+M_s)}{h \nu}
-\frac{(S+M_s-1)(S-M_s+2)}{h \nu - 2 D} \nonumber \\ 
& &-\frac{(S-M_s)(S+M_s+1)}{h \nu + 2 D}   \;.
\label{eq:hres_X} 
\end{eqnarray}
Notice that the sign of the term proportional to $X$
for the transition $M_s \rightarrow M_s-1$ is opposite to 
that for $M_s \rightarrow M_s+1$. 
The first type of transitions, $M_s \rightarrow M_s-1$, 
represent transitions among energy levels in the lower-energy
potential well (see Fig. \ref{energywell}). 
The second type, $M_s \rightarrow M_s+1$,
corresponds to transitions among energy levels in the higher-energy 
potential well (see Fig. \ref{energywell}).
At frequencies approximately lower than 100 GHz both types of transitions are 
observed, while
at higher frequencies only transitions in the lower energy are observed.
This is because at high frequencies the EPR excitation energy exceeds 
the difference in adjacent energy levels relevant to the transitions.

\begin{figure}
\begin{center}
\epsfxsize=7.5cm
\epsfysize=7.0cm
\epsfbox{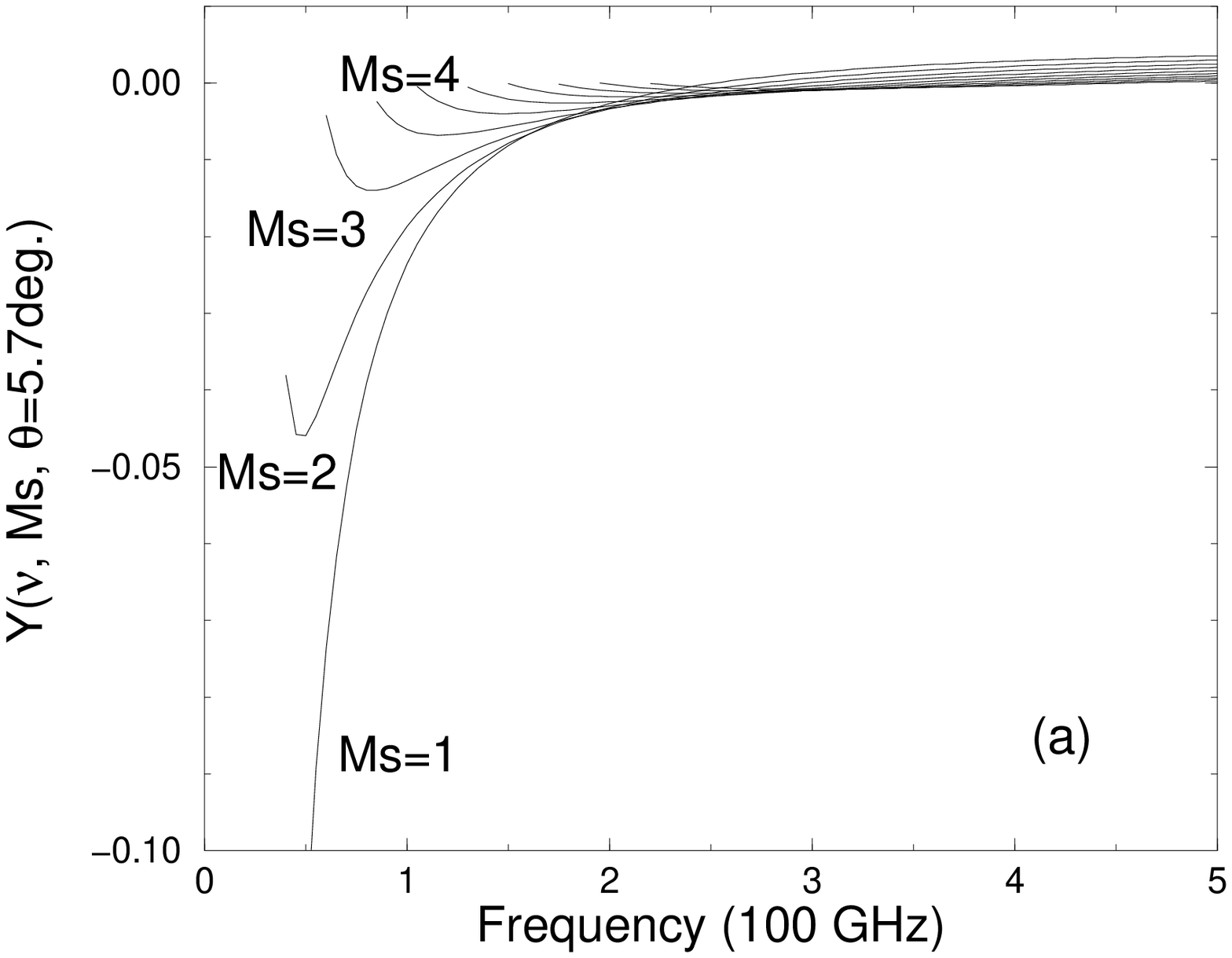}
\epsfxsize=7.5cm
\epsfysize=7.0cm
\epsfbox{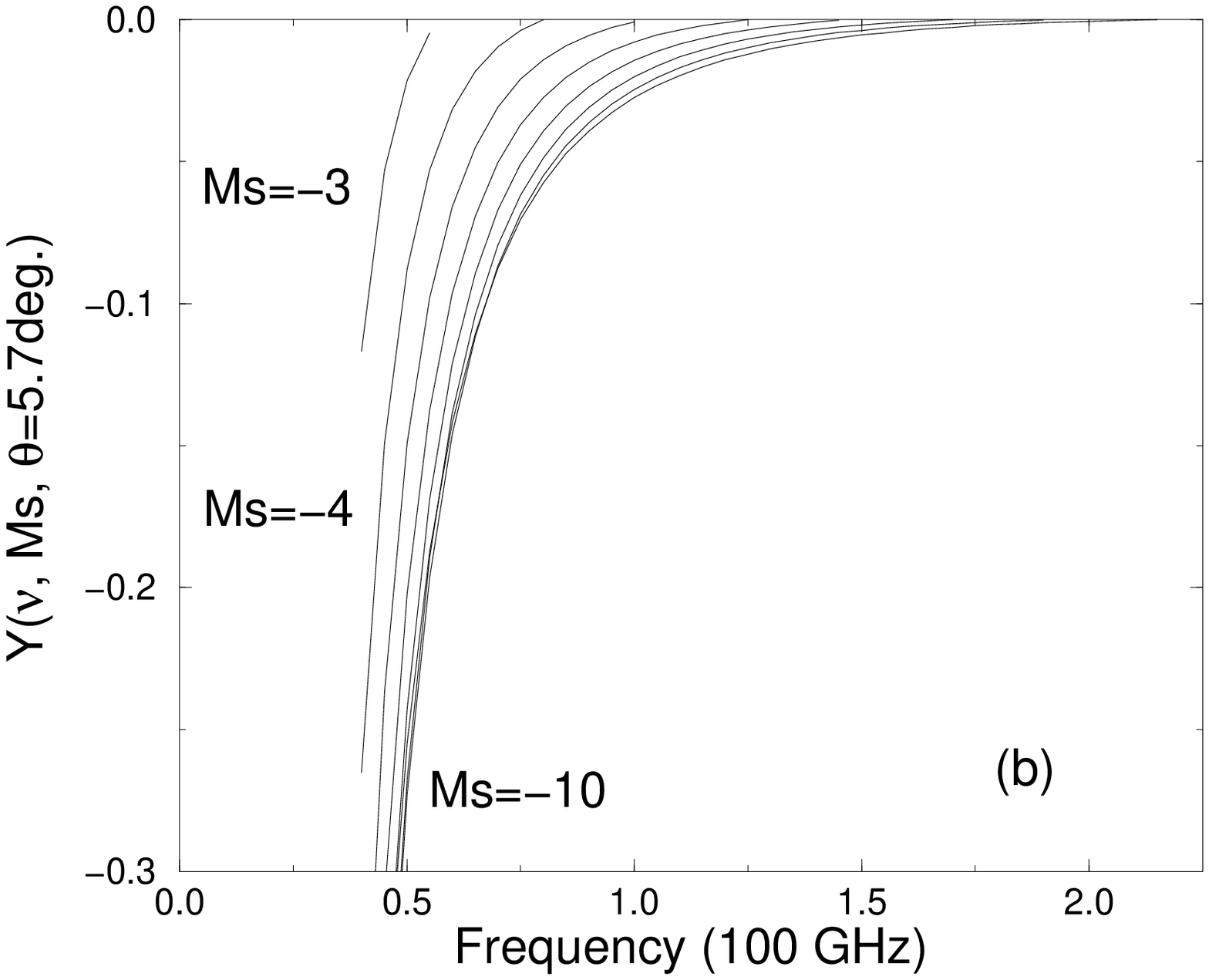}
\vspace{0.1in}
\caption{The ratio between the second-order and the zero-order
terms in the resonant fields, $Y(\nu,M_s,\theta)$, vs frequency
at $\theta \approx 5.7^{\circ}$ for the transitions 
(a) $M_s \rightarrow M_s-1$ and 
(b) $M_s \rightarrow M_s+1$. 
See details in the text.
Equations (\ref{eq:hres_hz_3}) and (\ref{eq:hres_hz_4}) are valid
when $|Y(\nu,M_s,\theta)| \ll 1$. In (a) $M_s=1$ represents
the transition $M_s=1 \rightarrow 0$. In (b) $M_s=-3$ 
represents the transition $M_s=-3 \rightarrow -2$.}
\label{fig:YYandYYM}
\end{center}
\end{figure}

Our approximation [Eqs.~(\ref{eq:hres_hz_3}) and (\ref{eq:hres_hz_4})]
is valid when the absolute value of 
the ratio between the second-order and the zero-order terms 
becomes very small compared to unity: 
$|Y(\nu, M_s, \theta)| \equiv |g \mu_B B_0
\sin^2\theta~X(\nu, \mp M_s)/(4 \cos\theta)| \ll 1$ for $M_s \rightarrow
M_s \pm 1$. Figure \ref{fig:YYandYYM} shows 
$Y(\nu, M_s, \theta)$ as a function of frequency $\nu$ and $M_s$ at 
$\theta=0.1$ rad ($\approx 5.7^{\circ}$) for the 
two types of transitions. For the transitions 
$M_s \rightarrow M_s-1$ (we consider the case that 
the resonant field is positive at $\theta=0$, that is,
$h\nu-D(2M_s-1) > 0$), the approximation becomes
good for all $M_s$ in the range of frequencies 
from 50 GHz to 500 GHz, except for higher energy levels
than $M_s=1$ [Fig. \ref{fig:YYandYYM}(a)]. 
The approximation is better for higher frequencies and {\it larger} $M_s$
(lower energy levels).
For the transitions $M_s \rightarrow M_s+1$ 
(we consider the case that $D(2M_s-1)-h\nu > 0$), the approximation
becomes better at higher frequencies or for {\it smaller} magnitude of $M_s$ 
(higher energy levels) [see Fig. \ref{fig:YYandYYM}(b)].
As shown in Fig. \ref{fig:Hres},
the resonant fields from this approximation agree well with
the exact diagonalization results at $\nu=66.135$~GHz.\cite{FREQ}

Next we examine the consequences of the nonzero $\theta$
on the resonant fields, Eqs.~(\ref{eq:hres_hz_3}) and (\ref{eq:hres_hz_4}),
as functions of the frequency and the energy level. 
Gaussian distributions
in $D$ and $g$ provide symmetric distributions in the resonant fields. 
However, a Gaussian distribution in $\theta$ lets the molecules
have smaller {\it or} larger resonant fields than that for $\theta=0$,
in an asymmetric fashion. This yields asymmetric tails in the 
average line shapes of EPR spectra. 

\begin{figure}
\begin{center}
\epsfxsize=8.0cm
\epsfysize=7.0cm
\epsfbox{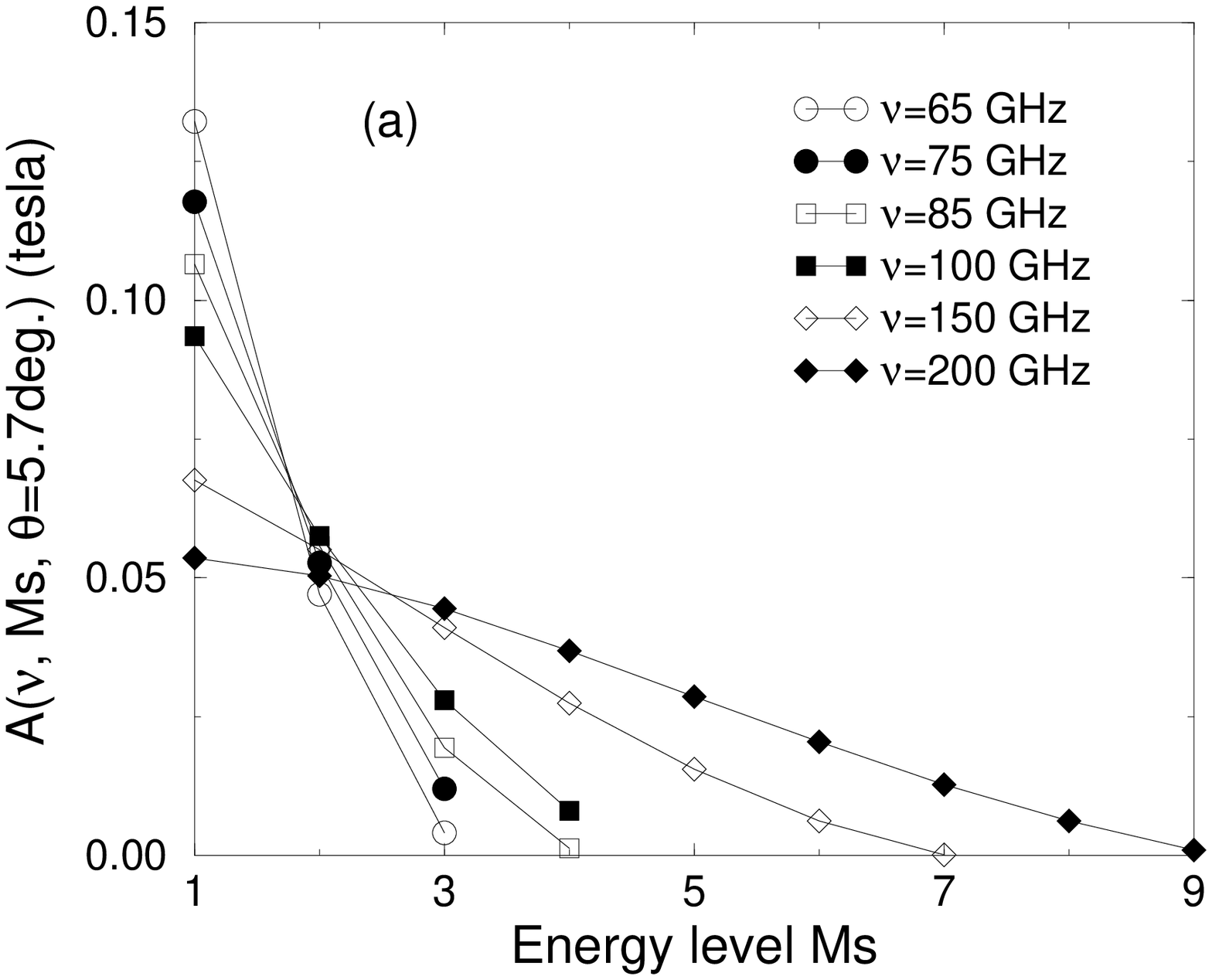}
\hspace{.2in}
\epsfxsize=8.0cm
\epsfysize=7.0cm
\epsfbox{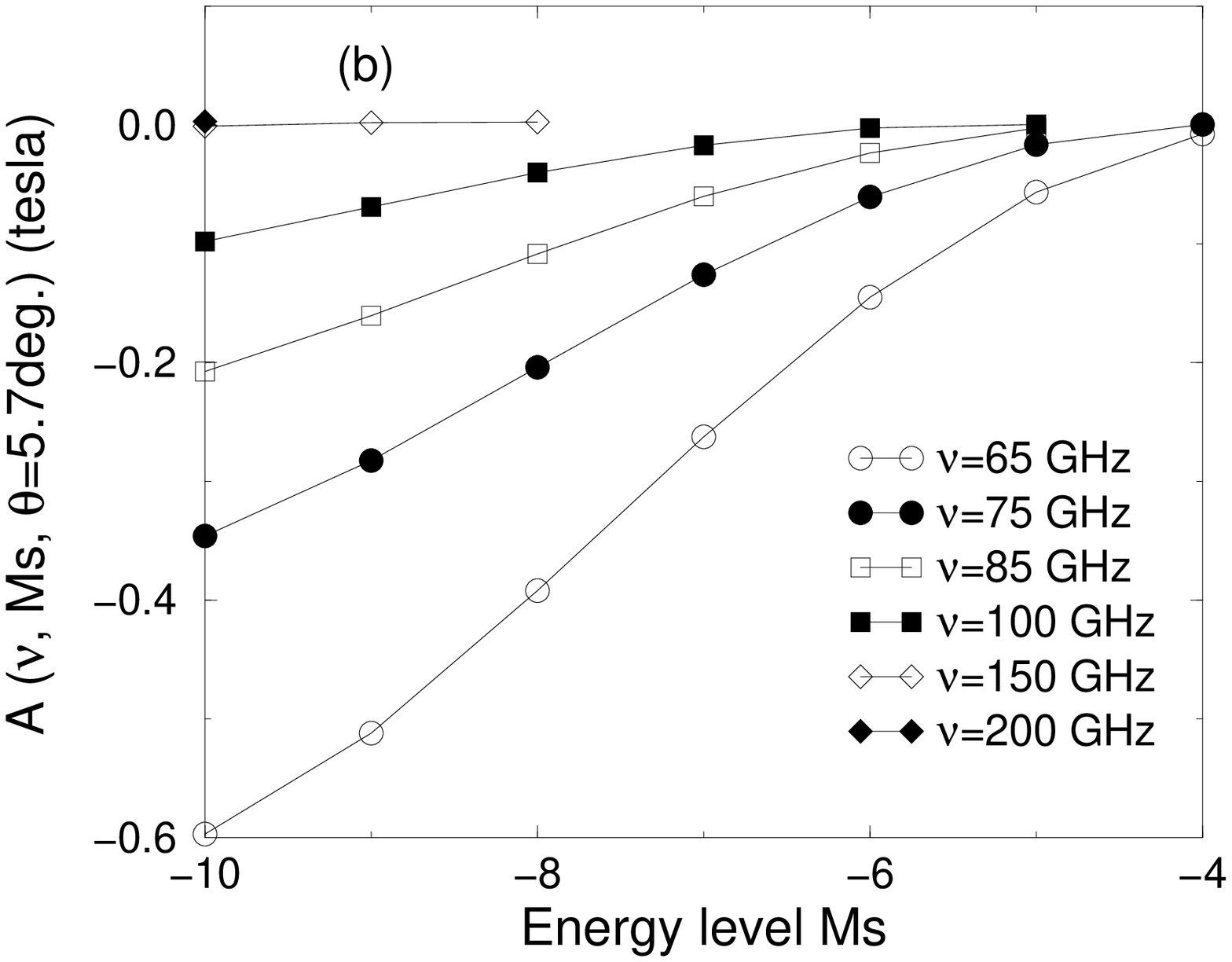}
\caption{Asymmetry $A$($\nu,M_s,\theta=5.7^{\circ}$) vs 
$M_s$ at various frequencies for the transitions (a)
$M_s \rightarrow M_s-1$ and (b) $M_s \rightarrow M_s+1$.
See details in the text.}
\label{fig:AvsMs}
\end{center}
\end{figure}

\begin{figure}
\begin{center}
\epsfxsize=8.0cm
\epsfysize=7.0cm
\epsfbox{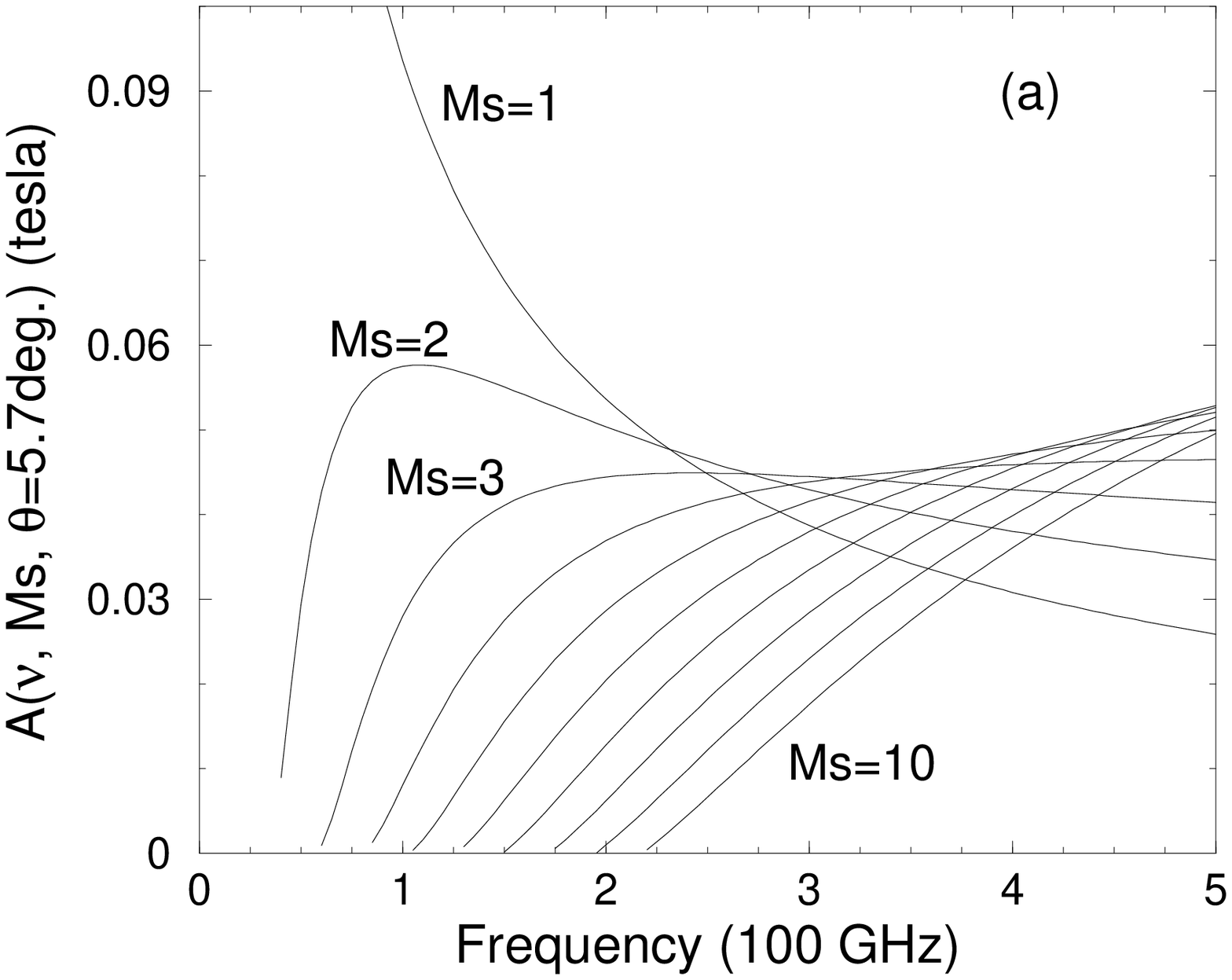}
\hspace{.2in}
\epsfxsize=8.0cm
\epsfysize=7.0cm
\epsfbox{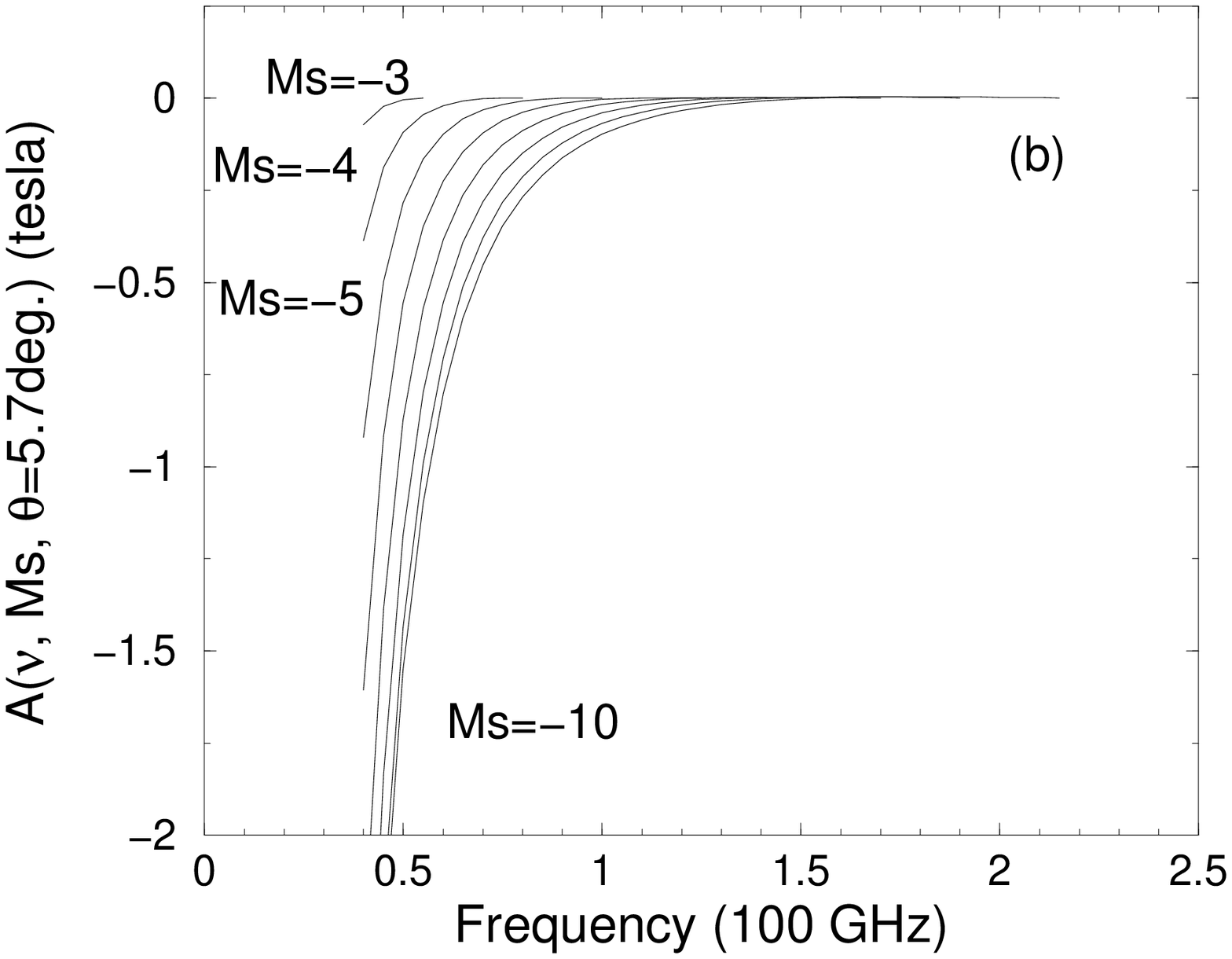}
\caption{Asymmetry $A$($\nu,M_s,\theta=5.7^{\circ}$) vs 
the frequency $\nu$ for the transitions (a)
$M_s \rightarrow M_s-1$ and (b) $M_s \rightarrow M_s+1$.
See details in the text.}
\label{fig:ASYMandASYMM}
\end{center}
\end{figure}

To investigate the asymmetry effect as a function of $\nu$ and 
$M_s$, we define the quantity $A(\nu,M_s,\theta)
\equiv B_{\mathrm res}(\theta\neq 0) - B_{\mathrm res}(\theta=0)$.
As shown in Figs.~\ref{fig:AvsMs}(a) and \ref{fig:ASYMandASYMM}(a), 
for $M_s \rightarrow M_s-1$, the sign of $A$($\nu,M_s$,
$\theta=5.7^{\circ}$)
is positive so the resonant fields become {\it minima} at $\theta=0$
for the whole examined range of frequencies and for all energy
levels in the lower-energy potential well. Therefore, the spectra will have
asymmetric tails in the direction of {\it increasing} field. The magnitude of 
$A$($\nu, M_s$, $\theta=5.7^{\circ}$) increases with 
decreasing $M_s$ at low frequencies (below about 200 GHz) 
[Fig. \ref{fig:AvsMs}(a)] because of high resonant fields, 
and decreases weakly with decreasing 
$M_s$ at high frequencies 
[Fig. \ref{fig:ASYMandASYMM}(a)]. 
For a particular $M_s$, the magnitude of
$A$($\nu, M_s$, $\theta=5.7^{\circ}$) increases with increasing 
frequency, except for $M_s=3 \rightarrow 2$, $M_s=2 \rightarrow 1$,
and $M_s=1 \rightarrow 0$ [Fig. \ref{fig:ASYMandASYMM}(a)]. 
For $M_s=3 \rightarrow 2$ and $M_s=2 \rightarrow 1$, 
the magnitude increases with frequency for low frequencies, 
and then starts to decrease at about 100 GHz and 250 GHz, respectively. 
For $M_s=1 \rightarrow 0$ the magnitude decreases monotonically
with increasing frequency.
As shown in Figs. \ref{fig:AvsMs}(b) and \ref{fig:ASYMandASYMM}(b), 
for $M_s \rightarrow M_s+1$, the sign of 
$A$($\nu,M_s$,$\theta=5.7^{\circ}$)
is negative so the resonant fields attain their {\it maxima} at $\theta=0$
for the whole examined range of frequencies and $M_s$ in the
higher-energy potential well.
Thus, the spectra show asymmetric tails in the direction 
of {\it decreasing} field.
The magnitude of $A$($\nu,M_s$, $\theta=5.7^{\circ}$)
increases with increasing magnitude of $M_s$ [Fig. \ref{fig:AvsMs} (b)] 
and with decreasing frequency [Fig. \ref{fig:ASYMandASYMM}(b)],
because the resonant fields increase.
It should be recalled that our approximation 
breaks down at low frequencies and large magnitude of
$M_s$ [Fig.~\ref{fig:YYandYYM}(b)]. 

Here we discuss some previous works using similar theoretical methods.
Bleaney\cite{BLEA51} used the similar spin Hamiltonian 
to Eq.~(\ref{eq:ham_hz_xyz}) in a strong field limit
to examine the hyperfine structure in paramagnetic salts.
Friedman and Low calculated the resonant fields 
for Mn$^{2+}$ doped into zinc fluosilicate (ZnSiF$_6$:6H$_2$O)
with the applied field oriented at an angle with respect to the crystal axes, 
assuming that the Zeeman energy is much larger than the zero-field 
anisotropy energy.\cite{FRIE60} This approximation, which is the same
as discussed in Appendix~A of the present paper, 
is a proper assumption for Mn$^{2+}$ in zinc 
fluosilicate since that system has a small crystal-field anisotropy.
If we apply this assumption to the single-molecule magnet Mn$_{12}$
however, we find that the resonant field within this approximation
is in poor agreement with the exact diagonalization results.
(See Fig. \ref{fig:Hres} and Appendix A.) The second-order corrections
overestimate the exact results. For Mn$_{12}$ at most intermediate
frequencies, the Zeeman energy is {\it not\/} very large compared to
the zero-field anisotropy, and 
the appropriate expressions for the resonant fields are our  
Eqs.~(\ref{eq:hres_hz_3}) and~(\ref{eq:hres_hz_4}).

\section{Distribution functions for the resonant fields}

\begin{figure}
\begin{center}
\epsfxsize=7.5cm
\epsfysize=7.0cm
\epsfbox{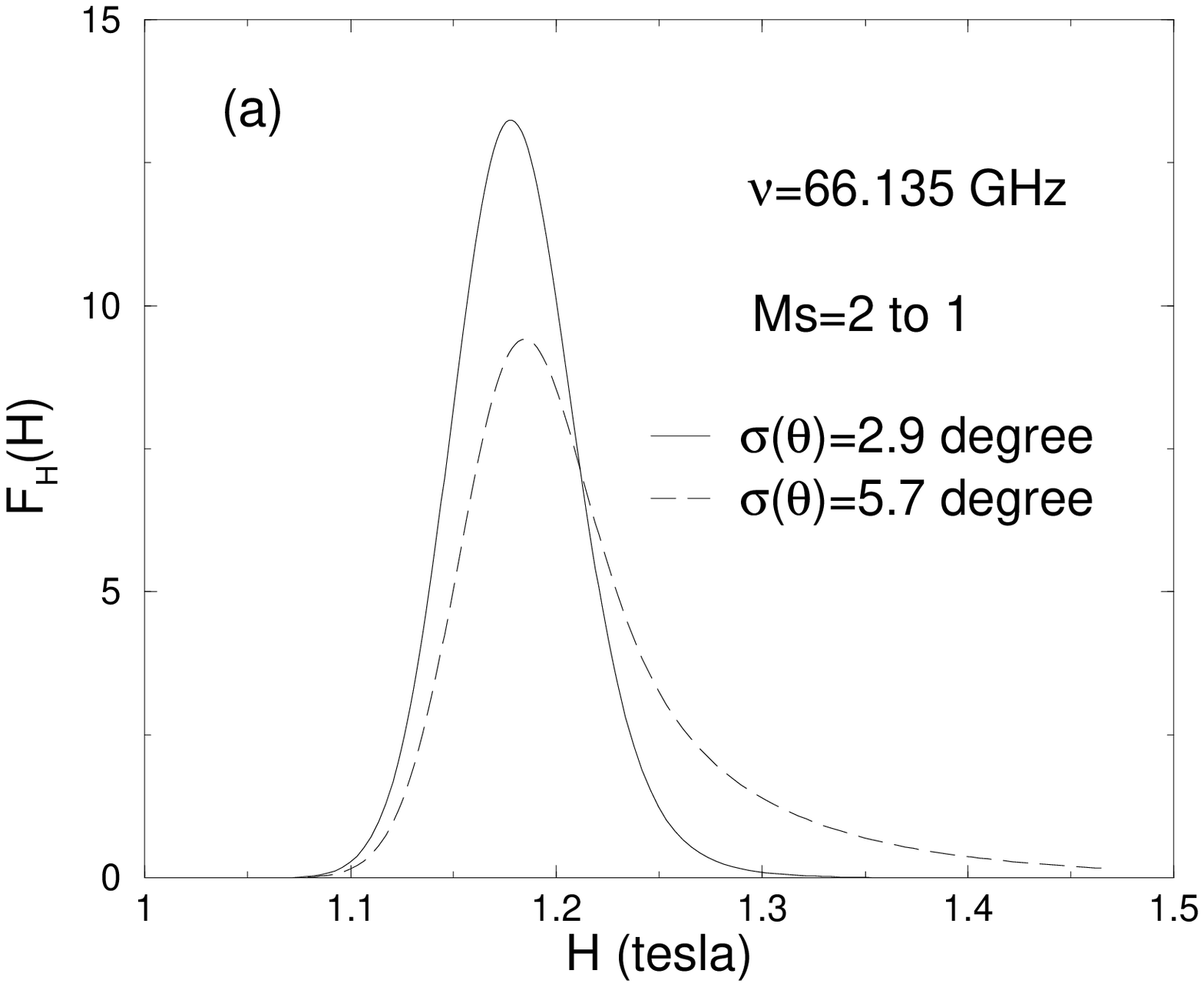}
\epsfxsize=7.5cm
\epsfysize=7.0cm
\epsfbox{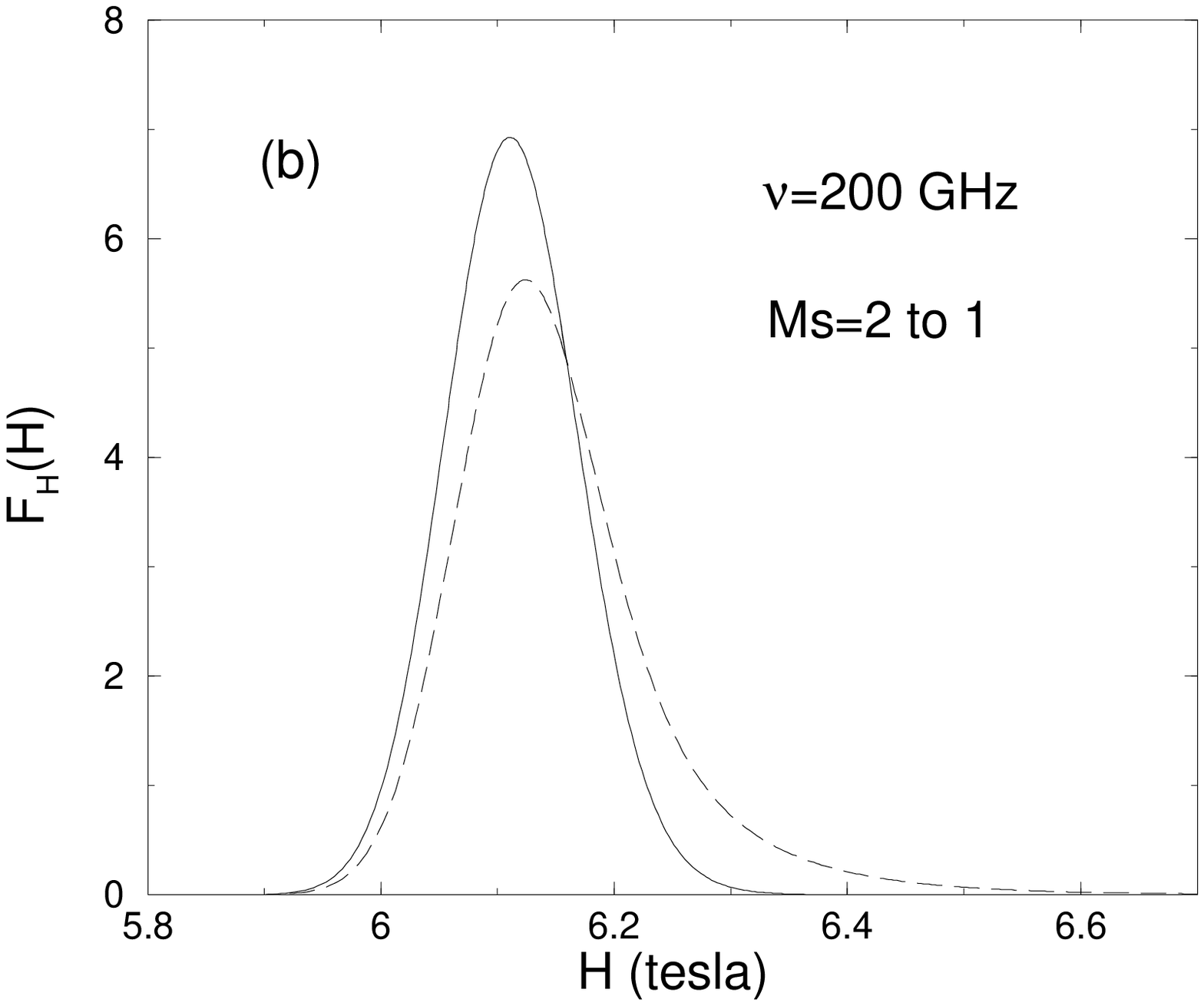}
\epsfxsize=7.5cm
\epsfysize=7.0cm
\epsfbox{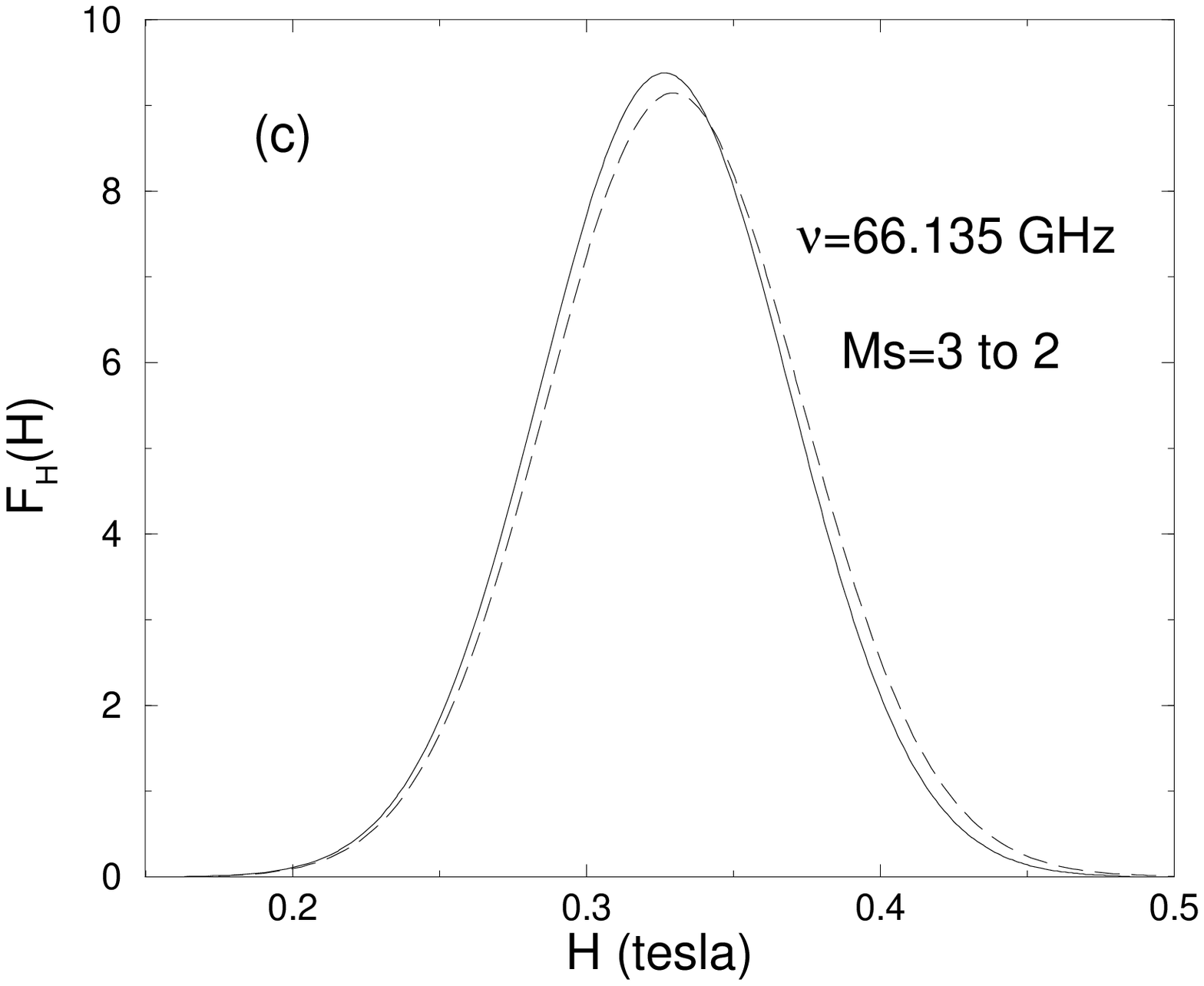}
\epsfxsize=7.5cm
\epsfysize=7.0cm
\epsfbox{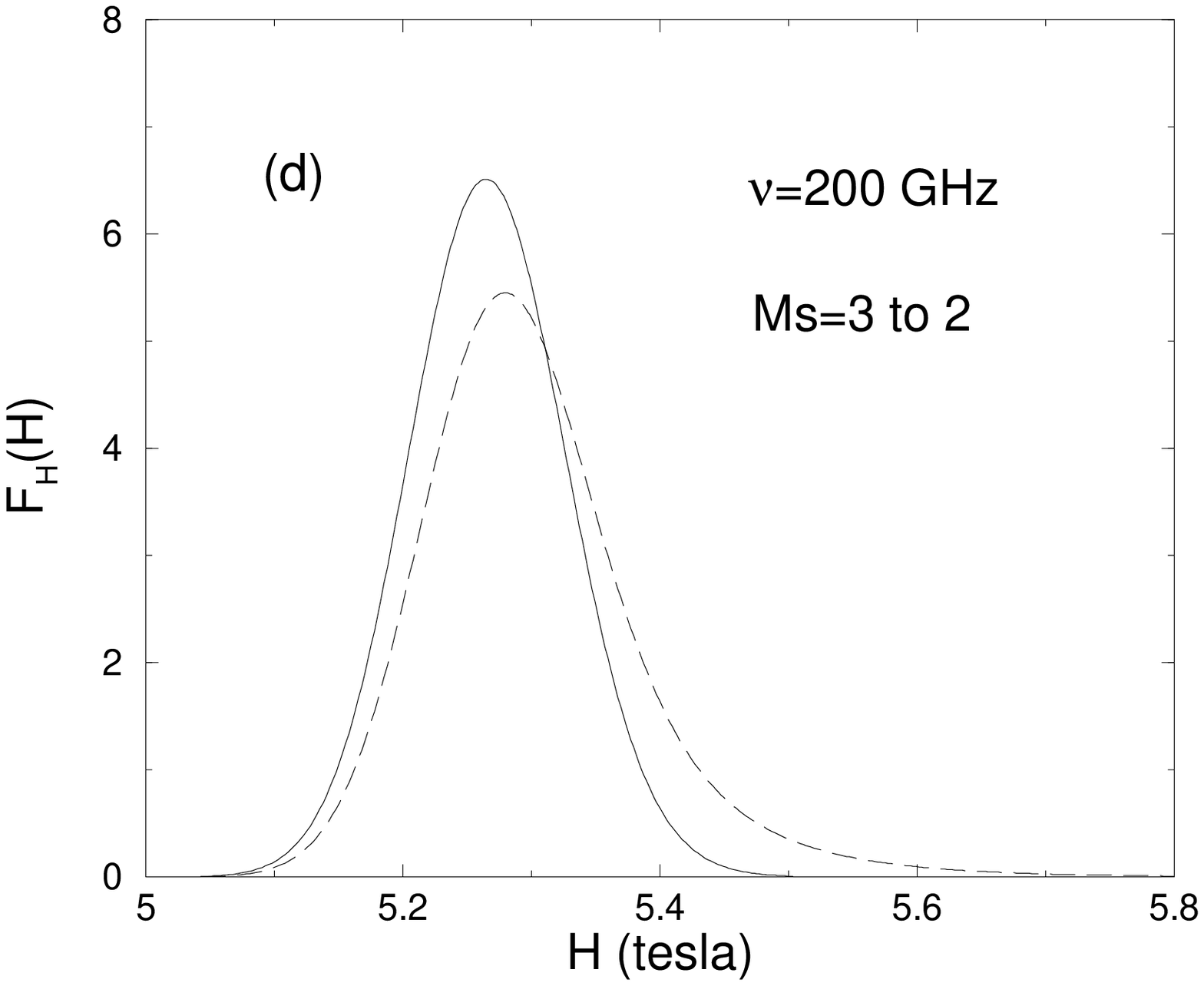}
\epsfxsize=7.5cm
\epsfysize=7.0cm
\epsfbox{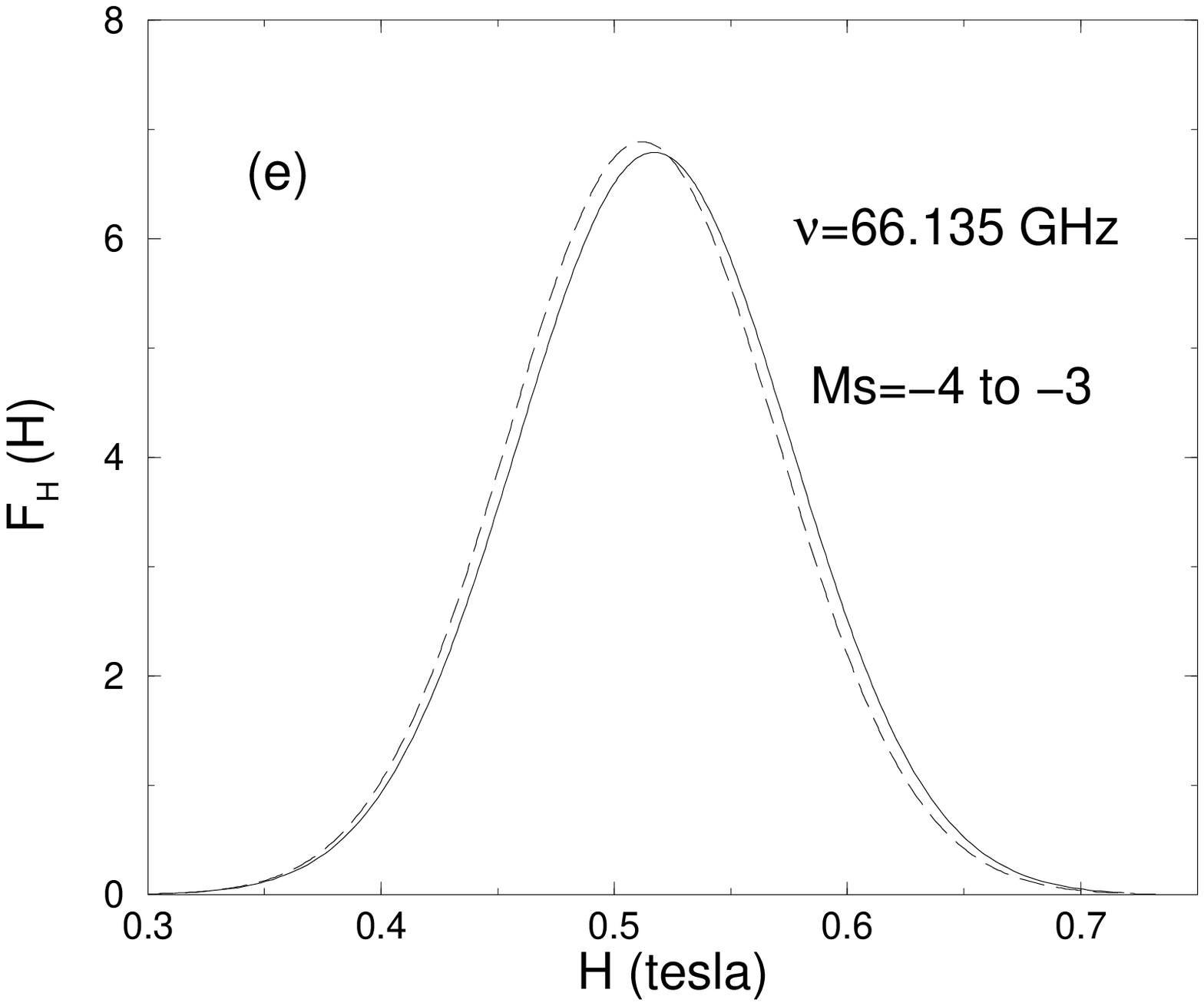}
\epsfxsize=7.5cm
\epsfysize=7.0cm
\epsfbox{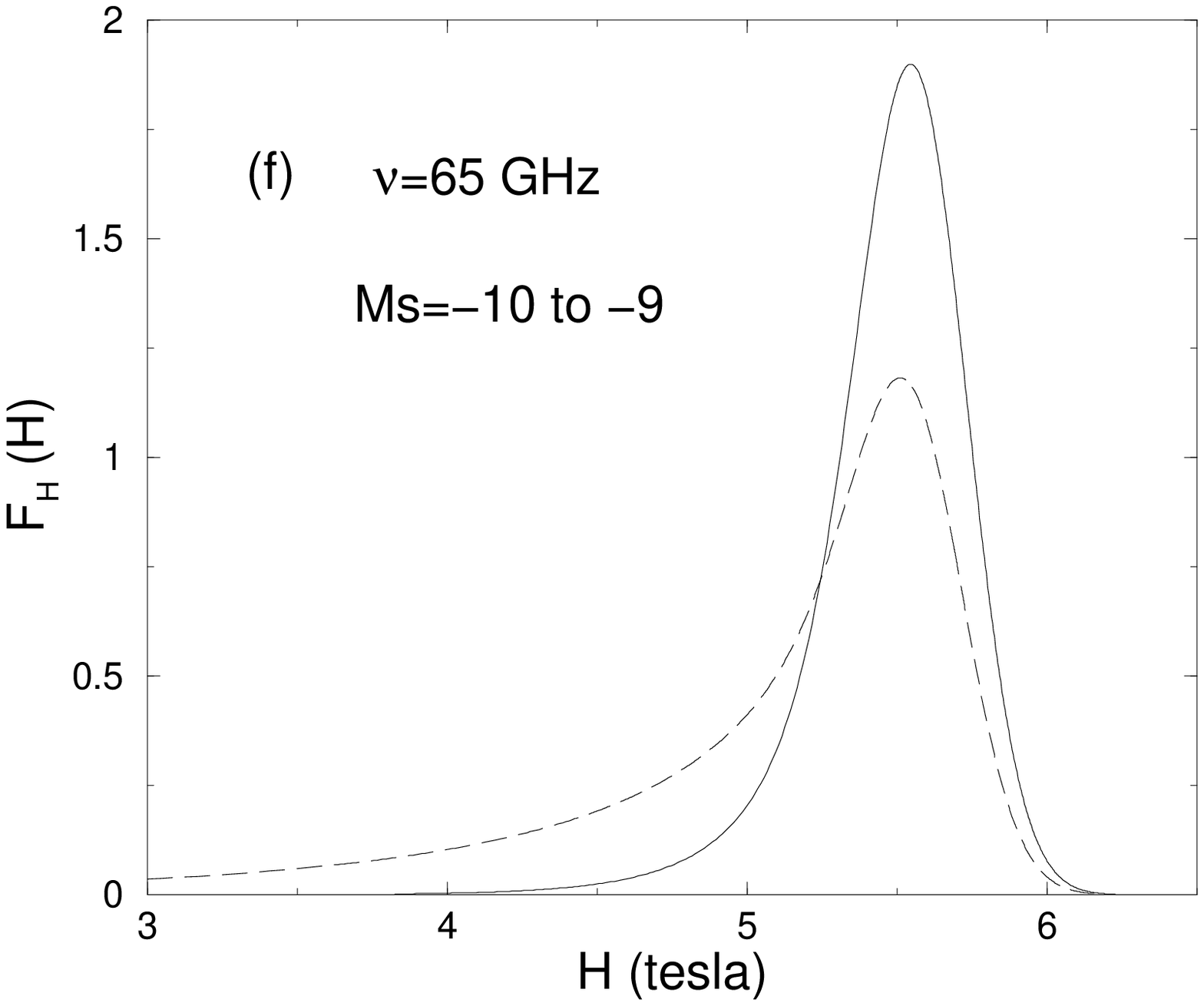}
\caption{The probability distribution functions for the resonant fields 
for the transitions: (a) $M_s=2\rightarrow 1$, 
(c) $M_s=3\rightarrow 2$, (e) $M_s=-4\rightarrow -3$ 
at $\nu=66.135~{\mathrm GHz}$, (b) $M_s=2\rightarrow 1$,
(d) $M_s$$=$3$\rightarrow 2$ at $\nu=200~{\mathrm GHz}$, 
and (f) $M_s=-10\rightarrow -9$ at $\nu=65~{\mathrm GHz}$, 
when $\vec{B} \| \hat{c}$.
Here we use the standard deviations of $D$ as 0.02$D$, the standard deviation
of $g$ as 0.008$g$, and the standard devation of $\theta$ as $2.9^{\circ}$
(solid curves) and $5.7^{\circ}$ (dashed curves). }
\label{fig:Fhhh}
\end{center}
\end{figure}

Since we obtained the resonant fields as functions of $g$, $D$, 
and $\theta$ in Eqs.~(\ref{eq:hres_hz_3}) and (\ref{eq:hres_hz_4}), 
we now calculate numerically the probability-distribution 
functions (pdf) for the resonant fields, using 
Gaussian distributions in $g$, $D$, and $\theta$, to predict the
experimental EPR spectra. In our current study, the asymmetry
in the spectra is concerned rather than the relative intensities
of the spectra. The effect of the perturbation $V$ due to
the nonzero $\theta$ (and the distribution in $\theta$)
on the relative intensities will be briefly discussed in Sec.~V.
The pdf of the resonant field, $F_{B}(B)$, 
can be calculated as follows.
\begin{eqnarray}
F_{B}(B)&=&\int d\theta \int d D \: f_{B,D,\theta} (B,D,\theta)\;, 
\label{eq:fabg} \\
f_{B,D,\theta} (B,D,\theta)&=&
f_{g,D,\theta}(g^{\ast}(B),D,\theta)/ 
\left|\frac{\partial B}{\partial g}(g^{\ast}(B),D,\theta) \right| \:, 
\end{eqnarray}
where $B$ denotes the resonant field, Eq.~(\ref{eq:hres_hz_3})
or (\ref{eq:hres_hz_4}), and $g^{\ast}(B)$ is obtained by 
solving the resonant field Eq.~(\ref{eq:hres_hz_3}) 
or (\ref{eq:hres_hz_4}) for $g$.\cite{PAPO} The function
$f_{g,D,\theta}(g^{\ast}(B),D,\theta)$ is the joint pdf of the
three random variables $g$, $D$, and $\theta$,
calculated at $g=g^{\ast}(B)$. 
For simplicity, we assume that 
$g$, $D$, and $\theta$ are statistically independent, so that
the joint pdf factorizes. The double integration in Eq.
(\ref{eq:fabg}) was performed numerically 
using {\texttt {Mathematica}}.\cite{MATH}
Figure \ref{fig:Fhhh} shows the pdfs of the resonant fields for
a few transitions at low ($\nu=65$ GHz and 66.135 GHz) and 
high ($\nu=200$ GHz) frequencies. Here we use
fixed values of the standard deviations 
of $D$ and $g$, ($\sigma_D=0.02D$ and $\sigma_g=0.008g$, which are
the same values used in Ref. \onlinecite{PARK02-1}) and two 
different values of the standard deviation of $\theta$, 
$\sigma_{\theta} \approx 2.9^{\circ}$ and $5.7^{\circ}$. 

As expected from the previous section (see Fig.~\ref{fig:ASYMandASYMM}), 
for $M_s=2 \rightarrow$ 1 [Figs. \ref{fig:Fhhh}(a) and (b) for $\nu=66.135$
GHz and $\nu=200$ GHz, respectively], 
the long tail of each pdf appears on the right hand side of the 
maximum of the pdf (the peak field). For both frequencies, the long 
tails are recognizable even for $\sigma_{\theta}\approx 2.9^{\circ}$, 
and they become prominent for $\sigma_{\theta}\approx 5.7^{\circ}$. 
For $M_s=3 \rightarrow$ 2, at $\nu=66.135$ GHz, 
it is hard to visually recognize any asymmetry 
[Fig. \ref{fig:Fhhh}(c)] because the asymmetry is much smaller 
and the symmetric linewidth is larger than 
for $M_s=2 \rightarrow$ 1. However, we see 
a small shift of the peak field towards higher fields as 
$\sigma_{\theta}$ increases from 2.9$^{\circ}$ to 5.7$^{\circ}$. 
As the frequency increases, the asymmetry effect
is significantly enhanced [Fig. \ref{fig:Fhhh}(d)]. 
For $M_s=-4 \rightarrow -3$, a possible asymmetry is 
expected at low frequencies only,
as seen from Fig.~\ref{fig:ASYMandASYMM}(b). The peak field 
shifts slightly towards lower fields as $\sigma_{\theta}$ 
increases from 2.9$^{\circ}$ to 5.7$^{\circ}$ [Fig. \ref{fig:Fhhh}(e)]. 
Other than that, it is hard to see any asymmetry in the pdf 
for the same reasons as for $M_s=3 \rightarrow$ 2.
For $M_s \rightarrow M_s+1$, as the magnitude of $M_s$ increases,
the asymmetry effect increases substantially
[Compare Fig.~\ref{fig:Fhhh}(e) with \ref{fig:Fhhh}(f)].
 
To quantify the asymmetry, we calculate the third central moment,
$\langle (B - \langle B \rangle)^3 \rangle$,
of the resonant-field distribution, and the difference, 
$(\langle B \rangle - B_{\mathrm peak})$,
between the average field and the peak field, 
for the transitions shown in
Fig. \ref{fig:Fhhh} (see Tables~\ref{table:asym1} and \ref{table:asym2}). 
If a pdf is symmetric, its third central moment vanishes, and the peak field 
should coincide with the average field. If a long tail appears on the 
right (left) hand side of the peak field, then the sign of the 
third central moment is positive (negative). 
For $M_s=2 \rightarrow$ 1 and $M_s=3 \rightarrow$ 2,
the signs of $\langle ( B - \langle B \rangle )^3 \rangle$
and $(\langle B \rangle - B_{\mathrm peak})$
are positive, whereas for $M_s=-4 \rightarrow -3$ and $M_s=-10 \rightarrow -9$
they are negative, in agreement with Fig.~\ref{fig:Fhhh}.

\section{Discussion and Conclusions}

As shown in Figs.~\ref{fig:AvsMs}-\ref{fig:Fhhh},
$\sigma_{\theta}\approx 5.7^{\circ}$ is large enough to observe
asymmetric tails in the EPR spectra for some frequencies and
energy levels, but small enough to be realistic in experimental samples.
We separately discuss the asymmetry in the spectra at a
particular value of $\sigma_{\theta}$ for 
$M_s \rightarrow M_s-1$ and for $M_s \rightarrow M_s+1$.
For $M_s \rightarrow M_s-1$ (in the lower-energy potential well), 
the asymmetry in the spectra
is more pronounced for smaller $M_s$ and higher frequencies, 
until the frequency becomes very high.
At very high frequencies (about 500 GHz or higher), 
the asymmetry in the resonant field increases weakly
with increasing $M_s$, but the asymmetry in the spectra for large $M_s$
would be masked by large symmetric linewidths. 
For $M_s \rightarrow M_s+1$ (in the higher well), the
asymmetry in the spectra is strong for large magnitude of $M_s$
and low frequencies, because of high resonant fields. 
But for very low frequencies (below about 50 GHz)
a different approximation scheme is needed than Eqs. (\ref{eq:hres_hz_3}) and 
(\ref{eq:hres_hz_4}). Since for high frequencies the EPR excitation 
energy becomes larger than the difference in adjacent energy levels,
the transitions $M_s \rightarrow M_s+1$ have narrow 
frequency or energy-level windows, in which asymmetric spectra are 
observable. Thus intermediate frequencies (between about 50 and 100 GHz)
would be optimum for observation of the two different types of
asymmetry in the spectra. 

So far we have concentrated on the asymmetry in the EPR line shapes
without mentioning intensities of the spectra.
Here we briefly discuss the effect of the distribution in the tilt angle 
$\theta$ on the intensities of the EPR spectra. At fixed values of
$D$, $g$, and $\theta$, the power absorbed between the $M_s$
and the $M_s-1$ energy level is written as
\begin{eqnarray}
\frac{d{\cal E}}{dt}
&=&{\cal E}_{M_s} [\dot{\rho}_{M_s,M_s}]
+  {\cal E}_{M_s-1} [\dot{\rho}_{M_s-1,M_s-1}],
\end{eqnarray}
where $\rho(t)$ is the density matrix of the spin system,
$\rho(t)_{m^{\prime} m}=\langle m^{\prime}| \rho(t) | m \rangle$,
$[\dot{\rho}_{M_s,M_s}]$ is the change with time
of the population in the $M_s$ level. To obtain $[\dot{\rho}_{M_s,M_s}]$,
the density matrix equation\cite{BLUM96} of the spin system 
[Eq.~(\ref{eq:ham_hz_xyz})] is used. For transition rates between
different energy levels, the spin-phonon coupling Hamiltonian\cite{LEUE00}
is used. For more technical details, refer to Ref.\onlinecite{PARK02-1,BLUM96}.
Near resonance (angular frequency of the external transverse 
oscillating magnetic field $B_x$, $\omega \sim ({\cal E}_{M_s} 
- {\cal E}_{M_s-1})/\hbar$),
the power absorption becomes \cite{BLUM96}
\begin{eqnarray}
\frac{d{\cal E}}{dt}  &=&
\frac{({\cal E}_{M_s-1}  -  {\cal E}_{M_s})}{\hbar^2}
|\langle M_s|B_x S_x|M_s-1 \rangle|^2 \:
\Delta(B_z) \: (\rho_{M_s,M_s}  - \rho_{M_s-1,M_s-1})~,
\label{eq:pwa}
\end{eqnarray}
where $\Delta(B_z)$ is a Lorentzian line-shape function\cite{PARK02-1}
and $\rho_{M_s,M_s}$ is the population of the $M_s$ level.
The average power absorption can be calculated by averaging
Eq.~(\ref{eq:pwa}) over the Gaussian distributions in $D$,
$g$, and $\theta$. The relative intensity of the average
power absorption at a fixed resonance frequency 
is determined by the transition probability,
the line-shape function, the population difference (Boltzmann
factors), and spreads in $D$, $g$, and $\theta$.  
The transition probability changes with the energy levels involved
with the EPR transition, but 
it does not change due to the perturbation $V$ 
caused by the nonzero $\theta$. The reason is that
as far as we are interested in the phenomena near resonance,
only perturbation terms which oscillate with the frequency 
close to the resonance frequency are important to the transition 
probability. Thus, the interaction, 
$Q(t)=B_x S_x (e^{i\omega t}+e^{-i\omega t})/2$, 
between the spin system and the oscillating transverse magnetic field 
$B_x$ does contribute to the transition probabilities, 
but not the perturbation terms $V$. Then, the maximum height of the
line-shape function $\Delta(B_z)$ is determined by a linewidth
due to the finite lifetime of an excited state. The lifetime of
an excited state does not change much with the perturbation terms $V$,
so that the height is not much affected by the nonzero $\theta$. 
The populations of the $M_s$ and $M_s-1$ level are not affected by 
the nonzero $\theta$, because the populations are determined by the
Boltzmann factor and small changes in the energy due to $V$ are
not noticeable in the Boltzmann factor.
Therefore, we conclude that the relative intensity for 
an EPR transiton between specific energy levels
is not affected by the perturbation $V$, and that
the effect of the distribution in $\theta$
on the relative intensity is 
solely due to the spread in $\theta$.

To observe asymmetric line shapes in EPR experiments,
one needs to optimize the experimental conditions.
First, the asymmetry
is more prominent for smaller $M_s$ for $M_s \rightarrow M_s-1$ (or
larger magnitude of $M_s$ for $M_s \rightarrow M_s+1$), 
but for those transitions 
the spectral intensity is generally poor at low temperatures 
because of the small populations in the excited states. 
Thus the sample temperature 
must be optimized to achieve a reasonable intensity.
Second, one must avoid the level crossings at which levels in the
two potential energy wells coincide.\cite{ZENE} If the
EPR transitions happen to occur near level-crossing points, 
the spectra could pick up a large extra line broadening which 
could prevent one from observing 
the small asymmetry. As the frequency increases, adjacent energy
levels move farther apart, so it becomes easier to avoid 
level-crossing points in the EPR transitions.
Third, one must avoid asymmetries caused by experimental artifacts,
such as the presence of standing waves in the EPR probe.
\cite{BARR97,BARR00} Standing waves can cause severe line-shape
distortion due to mixing between a
dispersion spectrum and an absorption spectrum. This
can be avoided by using the resonance cavity EPR technique.
\cite{HILL98,BARC00,BLIN01,MACC01} 
Fourth, the choice of frequency is important.
To observe two different types of asymmetry (one in the direction 
of increasing field and the other in the direction of decreasing field)
at a single frequency, the frequency must be lower than about 100 GHz.

If the asymmetries are observed in spectra with the optimum experimental
conditions, one can estimate
the distributions in $D$ and $g$ from the linewidths
of almost symmetric spectra. Then from the spectra
with significant asymmetries, we can estimate how broadly
the easy axes of the molecules are distributed in the sample.
The width of the distribution in $\theta$ may 
provide information on the defect concentration.
Additionally the distribution in $\theta$ may partially contribute
to the distribution in the tunnel splittings, which was 
recently proposed\cite{CHUD01} and measured experimentally,\cite{MERT01}
because nonzero $\theta$ produces the transverse terms relevant to
tunneling from one potential well to another.

In summary, we have theoretically examined the effect of defects 
on the line shapes of EPR spectra for field sweeps with
the quasi-static external field along the crystal $c$ axis at fixed
frequencies in the range of 50 GHz to 500 GHz
for a single crystal of the single-molecule magnet Mn$_{12}$. 
Static, random rotations of the magnetization easy axes of the molecules, 
due to defects, lead to asymmetries in the spectra. The strength and
direction of the
asymmetry depend on the frequency, the energy level, and the width
of the distribution in the tilt angle $\theta$ (which
depends on the defect concentration).
With carefully chosen frequencies, energy levels, and the appropriate
experimental technique, the asymmetry effect should be observable 
in future experiments. Comparison of the observed and calculated 
spectra could yield direct and quantitative information on the 
distribution in orientations and the concentration of defects
in the samples. This information may also be useful to understand the
origin of magnetization tunneling in the molecular
magnet Mn$_{12}$. Experimental examination of our predictions should be
quite fruitful.

\begin{center}
{\textbf{Acknowledgments}}
\end{center} 
We are grateful to M.~A.\ Novotny and S.\ Hill for
useful discussions.
This work was funded by NSF Grant Nos.~DMR-9871455, DMR-0120310,
and DMR-0103290, and by Florida State University through the School of 
Computational Science and Information Technology and
the Center for Materials Research and Technology. 

\appendix

\section{}

The spin Hamiltonian shown in this Appendix is equivalent to the 
spin Hamiltonian (\ref{eq:ham_hz_xyz}), but the approximation 
used in the calculations are different from those used in Sec. III. 

To rewrite this spin Hamiltonian in terms of the spin operators along 
the crystal axes, we use the following rotation matrix,\cite{GOLD}
\[ 
\left( \begin{array}{c}
S_x \\ S_y \\ S_z 
\end{array} \right) = 
\left( \begin{array}{ccc}
\cos\psi \cos\phi - \cos\theta \sin\phi \sin\psi & 
\cos\psi \sin\phi + \cos\theta \cos\phi \sin\psi &  
\sin\psi \sin\theta \\
- \sin\psi \cos\phi - \cos\theta \sin\phi \cos\psi & 
- \sin\psi \sin\phi + \cos\theta \cos\phi \cos\psi & 
\cos\psi \sin\theta \\
\sin\theta \sin\phi & -\sin\theta \cos\phi & \cos\theta 
\end{array}
\right)
\left( \begin{array}{c}
S_a \\ S_b \\ S_c
\end{array} \right)
\]
\label{eq:rotmtx}
where $(\theta, \phi, \psi)$ are the three Euler angles.\cite{GOLD}
Then we obtain
\begin{eqnarray}
{\cal H}&=& -D [ \sin^2\theta \sin^2\phi~S_a^2 
+ \sin^2\theta \cos^2\phi~S_b^2 + \cos^2\theta~S_c^2 
- \sin^2\theta \sin\phi \cos\phi~\{S_a,S_b \} \nonumber \\
& & -\sin\theta \cos\theta  \cos\phi~\{S_b,S_c \} 
+ \sin\theta \cos\theta \sin\phi~\{S_a,S_c \} ] - g \mu_B B S_c \:,
\label{eq:ham_hz_abc}
\end{eqnarray}
where $\{A,B \}$ is the anticommutator.

\begin{figure}
\begin{center}
\epsfxsize=7.5cm
\epsfysize=7.0cm
\epsfbox{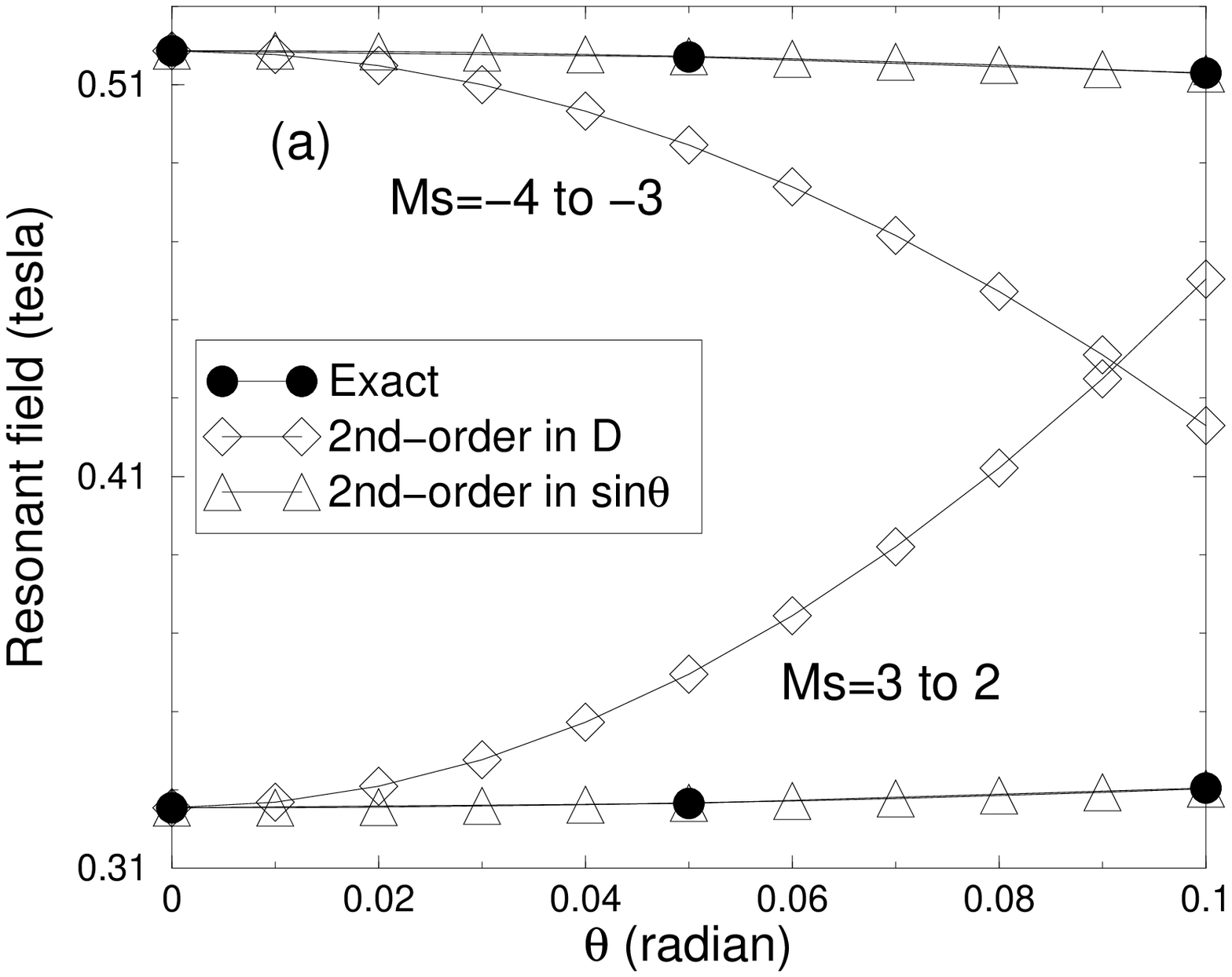}
\hspace{.2in}
\epsfxsize=7.5cm
\epsfysize=7.0cm
\epsfbox{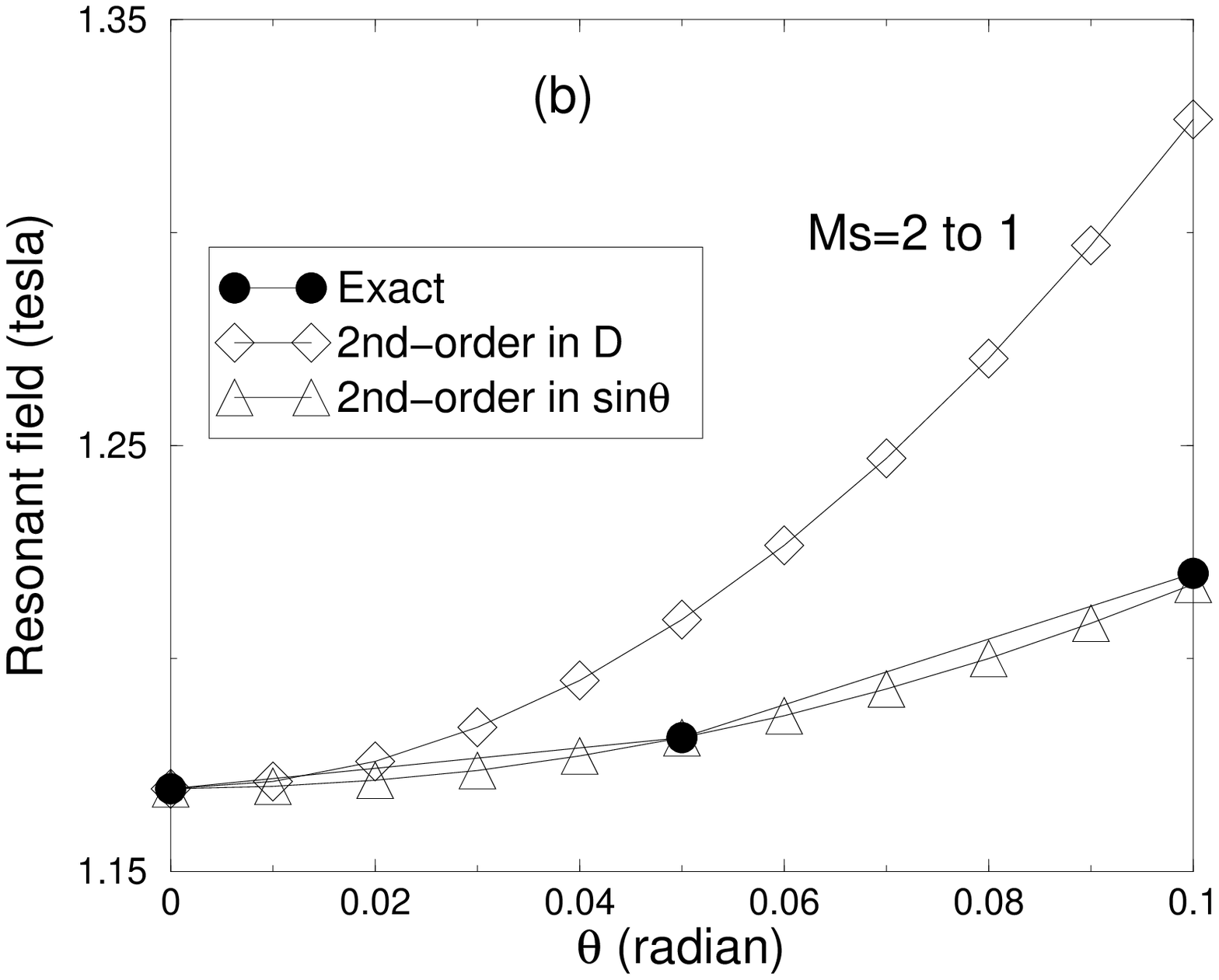}
\caption{Resonant fields vs tilt angle $\theta$
at $\nu=$66.135~GHz for the transitions (a) $M_s=3 \rightarrow 2$
and $M_s=-4 \rightarrow -3$, and (b) $M_s=2 \rightarrow 1$. 
The filled circles denote the exact results from numerical
diagonalization of the spin Hamiltonian. 
The triangles denote the results from Eqs. (\ref{eq:hres_hz_3}) and 
(\ref{eq:hres_hz_4}).
The diamonds denote the results from Eqs. (\ref{eq:hres_hz_1}) 
and (\ref{eq:hres_hz_2}). Eqs. (\ref{eq:hres_hz_3}) and (\ref{eq:hres_hz_4})
show good agreement with the exact results, 
in contrast to Eqs. (\ref{eq:hres_hz_1}) and (\ref{eq:hres_hz_2}).}
\label{fig:Hres}
\end{center}
\end{figure}

To calculate the resonant fields, we here assume that the Zeeman energy 
is much larger than the zero-field anisotropy energy, so that
the eigenvalues $M_s^{\prime}$ of $S_c$ are good quantum
numbers. Taking the Zeeman energy, $-g \mu_B B S_c$, as an unperturbed 
spin Hamiltonian (taking the quantization axis as the direction of
the external magnetic field) and the other terms as small perturbations, 
we obtain the resonant fields as follows:
\begin{eqnarray}
B_{\mathrm res}(M_s^{\prime} \rightarrow M_s^{\prime}-1)
&=& \frac{1}{g \mu_B} 
[ h \nu - \frac{2 M_s^{\prime} -1}{2} D (3 \cos^2\theta - 1) \nonumber \\
& & +\frac{D^2 \sin^2\theta \cos^2\theta}{2 h \nu} 
(431 +24 M_s^{\prime} -24 M_s^{\prime 2}) \nonumber \\
& & +\frac{D^2 \sin^4\theta}{8 h \nu} 
(-217 -6M_s^{\prime} +6M_s^{\prime 2}) ] 
\label{eq:hres_hz_1} \;, \\
B_{\mathrm res}(M_s^{\prime} \rightarrow M_s^{\prime}+1)
&=& \frac{1}{g \mu_B} 
[ -h \nu - \frac{2 M_s^{\prime}+1}{2} D (3 \cos^2\theta - 1) \nonumber \\
& &+\frac{D^2 \sin^2\theta \cos^2\theta}{2 h \nu} 
(-431 +24M_s^{\prime} +24M_s^{\prime 2}) \nonumber \\
& &+\frac{D^2 \sin^4\theta}{8 h \nu} 
(217 -6M_s^{\prime} -6M_s^{\prime 2}) ]\;.
\label{eq:hres_hz_2}
\end{eqnarray}
The above results were also shown in Ref.\onlinecite{FRIE60}.
As shown in Fig. \ref{fig:Hres}, the resonant fields in this
approximation do not agree well with the exact diagonalization
results, in contrast to Eqs.~(\ref{eq:hres_hz_3}) and
(\ref{eq:hres_hz_4}). For most intermediate
frequencies, the Zeeman energy is {\it not} very large
compared to the zero-field anisotropy energy, and the
approximation of Eqs.~(\ref{eq:hres_hz_3}) and
(\ref{eq:hres_hz_4}) is therefore to be preferred.


\begin{table}
\caption{Third central moments of the resonant-field distribution, and 
the difference between the average
and peak resonant fields at $\nu=65$ GHz, 66.135~GHz, and 200~GHz 
for $\sigma_{\theta} \approx 2.9^{\circ}$, $\sigma_D \approx 0.02 D$,
and $\sigma_g \approx 0.008 g$ (see the solid curves in Fig.~\ref{fig:Fhhh}). 
The third root is taken for the third central
moments to give the same units for both measures of the asymmetry.
For $M_s=-4$, the asymmetry is too small to obtain a reliable
moment and difference.}
\label{table:asym1}
\begin{tabular}{rrcc} 
frequency (GHz) & energy level $M_s$ & 
$\sqrt[3]{\langle (B- \langle B \rangle)^3 \rangle}$ (tesla) &
$\langle B \rangle - B_{\mathrm peak}$ (tesla) \\ \hline
66.135 & 2 & 0.023 & 
3.02 $\times 10^{-3}$ \\ \hline
66.135 & 3 & 0.0052  & 1.90 $\times 10^{-4}$ \\ \hline
65 & $-10$ &  $-0.27$ & $-6.68 \times 10^{-2}$ \\ \hline
200 & 2 & 0.030 & 3.07 $\times 10^{-3}$ \\ \hline
200 & 3 & 0.025 & 2.40 $\times 10^{-3}$ \\ 
\end{tabular}
\end{table}

\begin{table}
\caption{Third central moments of the resonant-field distribution, and 
the difference between the average
and peak resonant fields at $\nu=65$ GHz, 66.135~GHz, and 200~GHz 
with the same values of $\sigma_D$ and $\sigma_g$ as
in Table \ref{table:asym1}, but with
$\sigma_{\theta}=5.7^{\circ}$ (see dashed curves in Fig.~\ref{fig:Fhhh}).
}
\label{table:asym2}
\begin{tabular}{rrcc} 
frequency (GHz) & energy level $M_s$ & 
$\sqrt[3]{\langle (B- \langle B \rangle)^3 \rangle}$ (tesla) &
$\langle B \rangle - B_{\mathrm peak}$ (tesla) \\ \hline
66.135 & 2 & 0.068 & 
9.89 $\times 10^{-3}$ \\ \hline
66.135 & 3 & 0.016 & 1.11 $\times 10^{-3}$ \\ \hline
65 & $-10$ &  $-0.93$ & $-4.46 \times 10^{-1}$ \\ \hline
200 & 2 & 0.098 & 2.69 $\times 10^{-2}$ \\ \hline
200 & 3 & 0.091 & 2.13 $\times 10^{-2}$ \\ 
\end{tabular}
\end{table}

\begin{references}
\bibitem{FREQ} The frequency $\nu=66.135$ GHz corresponds to 
that used in Ref. \onlinecite{HILL98}.
\end{references}
\end{document}